\documentclass[prl,twocolumn]{revtex4-1}
  \usepackage{graphicx}
  \usepackage{amsmath}
  \usepackage{amssymb}
  \usepackage{makeidx}
  \usepackage{amsfonts}
  \usepackage{appendix}
  \usepackage{hyperref}
  \usepackage{mathrsfs}
  
  \hypersetup
  {
    colorlinks,%
    citecolor=black,%
    linkcolor=black,%
    urlcolor=black,%
  }

  \newcommand{\ket}[1]{\left| #1 \right\rangle}

\makeindex

\begin{document}

\title {Necessary and Sufficient Criteria for Gaussian Quantum Teleportation}

\author{Soumyakanti Bose}
\email{soumyabose@iisermohali.ac.in}
\affiliation{Indian Institute of Science Education and Research \\ 
Knowledge city, Sector 81, Sahibzada Ajit Singh Nagar, Punjab 140306, India.}

\date{\today}

\begin{abstract}
Quantum teleportation (QT) serves as one of the building blocks of the current state information science and technology which necessitates proper characterization of the resources yielding QT.
While entanglement is known to be the basic requirement to achieve QT, condition of sufficiency for QT still remains an open question.
Here, we provide an analytic proof that in the case of Gaussian entangled resources, in general, Einstein-Podolsky-Rosen (EPR) correlation is {\em sufficient for QT}.
For a relatively restricted set of Gaussian states we provide even a tighter condition that EPR correlation is {\em both necessary and sufficient for QT}.
Our results, in turn, provide a complete characterization of the resources required for QT.
\end{abstract}

\keywords{PACS: 03.67 Bg, 03.67 Mn, 42.50 Ex}

\maketitle

{\em Introduction:} 
Quantum teleportation (QT) plays central role in various information processing tasks, from broadband communication \cite{teleport_broadband} to quantum computing \cite{teleport_quantcomp}, quantum network \cite{teleport_network} to secret key distillation \cite{teleport_qkd} {\em etc}.
Resources yielding QT are the entangled states which could be realized in terms of discrete spin systems as well as continuous mode optical systems \cite{qit_book_nielsen, qit_book_wilde}.
After its first description by Bennett {\em et. al.} \cite{dvtp_bennett} for the spin systems, its extension to the quantum optical systems was proposed by Braunstein and Kimble (BK) \cite{cvtp_braunstein_kimble} and subsequently realized in various experimental set-ups\cite{tp_cv_exp1, tp_cv_exp2, tp_cv_exp3, tp_cv_exp4, tp_cv_exp5, tp_cv_exp6, tp_cv_exp7}. 

Recent analysis of both Gaussian and non-Gaussian resource states have pointed out various attributes such as Einstein-Podolsky-Rosen (EPR) correlation, Hillery-Zubairy (HZ) correlation \cite{epr_yang, epr_lee, epr_wang, hz_hu} of the resource states playing crucial role in achieving quantum fidelity.
Moreover, based on the no-cloning theorem \cite{nocloning_cerf}, People have also proposed quantum steering as the necessary condition for secure teleportation (ST) \cite{st_grosshans, st_he} - a much stricter condition than QT.
No-cloning bound has also been shown to play crucial role in teleporting input nonclassicality \cite{teleport_nonclassicality}.

For the most commonly used entangled Gaussian resource in QT - the two-mode squeezed vacuum state (TMSV) that belongs to a much restricted subclass of all Gaussian states, namely the symmetric states, entanglement turns out to be both {\em necessary and sufficient} for QT \cite{adesso_telport_symgauss}.
However, the same doesn't hold true in the case of general Gaussian entangled resources which form a much larger set than the set of symmetric states.
Consequently, in the context of teleportation based Gaussian information processing \cite{Weedbrook_Gauss_Quant_Inf}, in general, besides entanglement as the primary requirement, {\em sufficiency for QT is still an open question}.

In this letter, we resolve the issue of sufficiency condition (s) for QT with Gaussian entangled resources, in general.
We provide an analytic proof that the Einstein-Podolsky-Rosen (EPR) correlation is sufficient to ensure QT with entire set of Gaussian resources - in the sense that states having EPR correlation necessarily yields QT.
Moreover, for a restricted set of Gaussian states that is yet sufficiently bigger than the set of symmetric states, we obtain a even a tighter condition on the resource suitable for QT.
We prove that, in the case of all Gaussian resources for which the diagonal elements of the  correlation matrix (discussed in the text) are exactly opposite to each other, EPR correlation turns out to be both {\em necessary and sufficient} for QT.
Within the same framework, we also retrieve the earlier result that entanglement alone is necessary and sufficient QT with symmetric Gaussian resources.
This, in conjunction with the secure teleportation \cite{st_grosshans, st_he}, provides a complete characterization of Gaussian QT that could be immediately extended to other teleportation based information processing tasks.

We elaborate our results by considering the examples of two-mode squeezed thermal state (TMST) and beam splitter (BS) generated Gaussian states, as resource.
In the case of TMST that belongs to the restricted class of Gaussian states as mentioned above where the individual mode thermal parameters are different, we show that EPR correlation is a stricter condition than entanglement for QT.
While entanglement is necessary, EPR correlation is both necessary and sufficient for QT.
Moreover, fidelity of teleportation with TMST as resource turns out to be a monotonic function of the degree of EPR correlation.
We also note that with increase in squeeze parameter the difference between entanglement and EPR correlation reduces due to the increasing symmetry in the state.
On the other hand, BS output Gaussian resources, generated from input squeezed thermal state at one of inputs while other input left at vacuum, belong to the most general asymmetric Gaussian states.
Such resources provide example where EPR correlation is only sufficient, not necessary for QT.
Unlike the previous two mode case, here with increase in squeeze parameter the asymmetry of the state increases leading to more vivid difference between EPR correlation and QT.

{\em Gaussian States and Quantum Teleportation:}
A bipartite Gaussian state $\rho_{\rm{ab}}^{\rm{g}}$ is uniquely defined in terms of its covariance matrix $V_{\rm{ab}}$ defined as $V_{\rm{ab}} = \frac{1}{2}\rm{Tr} \left[ \rho_{\rm{ab}}^{\rm{g}} \lbrace \Delta {\bf R}, \Delta {\bf R}^{T} \rbrace \right]$.
The column matrix ${\bf R}$ is the vector whose elements are the quadrature components corresponding to modes $A$ and $B$ having the form ${\bf R} = (x_{a},p_{a},x_{b},p_{b})^{T}$ and $\Delta {\bf R} = {\bf R} - \rm{Tr} [\rho_{\rm{ab}}^{\rm{g}} {\bf R}]$.
The $4\times 4$ symmetric matrix $V_{\rm{ab}}$ satisfies the canonical uncertainty relation \cite{gs_duan, gs_simon, gs_serafini}
\begin{equation}
V_{\rm{ab}} + \frac{i}{2}\Omega \geq \frac{1}{2} ~~;~~ \Omega = \begin{pmatrix}
J & 0\\
0 & J
\end{pmatrix},
\label{eq_exp_cur}
\end{equation} 
where $J=\begin{pmatrix}
0 & 1\\
-1 & 0
\end{pmatrix}$.
It is well known that any bona-fide quantum covariance matrix for any inseparable state $\rho_{\rm{ab}}^{\rm{g}}$, by using proper symplectic transformations, could be brought into the canonical form
\begin{equation}
V_{\rm{ab}} = 
\begin{pmatrix}
\eta & 0 & c_{1} & 0 \\
0 & \eta & 0 & -c_{2} \\
c_{1} & 0 & \zeta & 0\\
0 & -c_{2} & 0 & \zeta
\end{pmatrix} = 
\begin{pmatrix}
A & C \\
C^{T} & B
\end{pmatrix}
\label{eq_exp_tmvm_canonical}
\end{equation}
where $A = \rm{diag} (\eta, \eta)$ and $B = \rm{diag} (\zeta, \zeta)$ are the covariance matrix corresponding to the each subsystems and $C = \rm{diag} (c_{1}, -c_{2})$ is the correlation matrix.
The opposite sign of $c_{1}$ and $c_{2}$ is considered in line with the condition of inseparability ({\em "entanglement"}) \cite{gs_simon}.
Throughout the paper we work with this canonical form and its various ramifications under various assumptions to obtain specific results.

In the Braunstein-Kimble (BK) \cite{cvtp_braunstein_kimble} protocol, the performance/success of teleportation is measured in terms of the fidelity of teleportation ($F$), defined as the overlap between the unknown input state and the output state (the retrieved state), $F = \rm{Tr} [\rho_{\rm{in}}\rho_{\rm_{out}}]$. 
The evaluation $F$ becomes particularly simple in the characteristic function (CF) description \cite{TPF_CF}. 
The CF of an $n$-mode quantum optical state $\rho$ is defined as $\chi_{\rho}(\lbrace \lambda_{i} \rbrace) = \rm{Tr} [\rho D(\lbrace \lambda_{i} \rbrace)]$ where $D(\lbrace \lambda_{i} \rbrace)=\Pi_{i=1}^{n}\exp [\lambda_{i} a^{\dagger}_{i}-\lambda^{*}_{i}a_{i}]$ is the $n$-mode displacement operator and $a_{i}$ is the $i^{\rm{th}}$-mode annihilation operator.
For any two-mode state $\rho_{\rm{ab}}$ as a resource, the fidelity of teleportation of an unknown input state $\rho_{\rm{in}}$ can be expressed as,
\begin{equation}
F=\int \frac{d^{2}\lambda}{\pi}~ \chi_{\rm{in}}(-\lambda)~\chi_{\rm{in}}(\lambda)~\chi_{\rm{ab}}(\lambda,\lambda^{*}),
\label{eq_exp_telfid_cf}
\end{equation}
where, $\chi_{\rm{in}}(\lambda)$ and $\chi_{\rm{ab}}(\lambda,\lambda^{*})$ are the CFs of $\rho_{\rm{in}}$ and $\rho_{\rm{ab}}$ respectively. 
In the case of a coherent state $\ket \alpha$ taken as the unknown input state, Eq. (\ref{eq_exp_telfid_cf}) reduces to
\begin{equation}
F = \int \frac{d^{2}\lambda}{\pi}~ e^{-\lambda^{2}} ~\chi_{\rm{ab}}(\lambda,\lambda^{*}).
\label{eq_exp_telfid_cs_cf}
\end{equation}

The maximum fidelity of teleportation of a coherent state attainable by a separable state in the BK protocol is $1/2$ \cite{TPF_CS}. 
Hence, QT is described in terms of Eq. (\ref{eq_exp_telfid_cs_cf}) as $F>1/2$.
In the case of the Gaussian state with covariance matrix given in Eq. (\ref{eq_exp_tmvm_canonical}) teleportation fidelity of a coherent state is given by \cite{Pirandola_LasPhys},
$F=1/\sqrt{\det[\mathscr{M}]}$,
where $\mathscr{M}=A - \lbrace \sigma_{z}, C \rbrace + \sigma_{z} B \sigma_{z} + I$, $\sigma_{z} = \rm{diag}(1,-1)$ is the Pauli spin matrix and $I$ stands for $2\times 2$ identity matrix.
Evidently, the condition of QT ($F>1/2$) boils down to
\begin{align}
\det [\mathscr{M}] &= 1 + 4c_{1}c_{2} + ((\eta + \zeta) + 2)((\eta  + \zeta) - (c_{1} + c_{2})) 
\nonumber
\\ 
&~~~~~~~~~~~~~~~~~~~~~~ - (\eta + \zeta)(c_{1} + c_{2}) < 4
\label{eq_cond_qt_gvm}
\end{align}

{\em EPR Correlation is Sufficient for QT:}
EPR correlation \cite{gs_duan} is defined in terms of the EPR uncertainty $\Delta_{\rm{EPR}}$ which is nothing but the sum uncertainty of the correlated {\em positions} $(x_{a} - x_{b})$ and anti-correlated {\em momenta} ($p_{a} + p_{b}$).
A Gaussian state $\rho_{\rm{ab}}^{\rm{g}}$ is said to be EPR correlated if the EPR uncertainty becomes smaller than $2$, i.e.,
\begin{align}
\Delta_{\rm{EPR}} &= \left\langle \Delta^{2} \left( x_{a} - x_{b} \right) \right\rangle + \left\langle \Delta^{2} \left( p_{a} + p_{b} \right) \right\rangle 
\nonumber
\\
&= \rm{Tr} [A] + \rm{Tr} [B] - \rm{Tr} [\lbrace \sigma_{z},C \rbrace]
\nonumber
\\
&= 2 \left( (\eta + \zeta) - (c_{1} + c_{2}) \right) < 2
\nonumber
\\
\Rightarrow &~ (\eta + \zeta) - (c_{1} + c_{2}) < 1,
\label{eq_cond_eprc_gvm}
\end{align}

Lesser the $\Delta_{\rm{EPR}}$ is than $2$ more correlated the state is.
In fact, one may further consider the degree of EPR correlation, $f_{\rm{EPR}} = \max \left( 0,2 - \Delta_{\rm{EPR}} \right)$, that plays an important role in QT with quantum optical resources \cite{epr_yang, epr_lee, epr_wang, hz_hu}.
$f_{\rm{EPR}}$ is chosen in way to ensure that it is zero for $\Delta_{\rm{EPR}}>2$.

Let's first consider that the state $\rho_{\rm{ab}}^{\rm{g}}$ is EPR correlated, i.e.,
$(\eta + \zeta) - (c_{1} + c_{2}) = 1-\epsilon ~;~ 0< \epsilon \leq 1$.
A straightforward calculation leads to 
\begin{align}
\det [\mathscr{M}] &= 1 + 4c_{1}c_{2} + ((\eta + \zeta) + 2)((\eta  + \zeta) - (c_{1} + c_{2})) 
\nonumber
\\ 
&~~~~~~~~~~~~~~~~~~~~~~~~~~~~~~~~~~~~~ - (\eta + \zeta)(c_{1} + c_{2})
\nonumber
\\
&= 4 - \epsilon(4 - \epsilon) - (c_{1} - c_{2})^{2} < 4 ~ \forall~ 0< \epsilon \leq 1.
\label{eq_cond_epr_tf_gvm}
\end{align}

Since the last term is always positive, for all $\epsilon>0$ we have $\det [\mathscr{M}]<4$ which indicates that {\em EPR correlation is sufficient for QT}.
One can further note from Eq. (\ref{eq_cond_epr_tf_gvm}) that with increase in $\epsilon$, $\det [\mathscr{M}]$ decreases below $4$ and thus the quantum fidelity increases. 
Consequently, for any Gaussian entangled state with EPR correlation, fidelity of QT is directly proportional its degree of EPR correlation ($f_{\rm{EPR}}$).

{\em EPR Correlation is necessary and sufficient for a restricted set of states:} Let's now ask the opposite question, i.e., whether EPR correlation is necessary for QT.
It is, in general, difficult to answer since EPR correlation is sufficient for inseparability implying if the state is EPR correlated it is definitely entangled while the converse is not true.
To answer the question, we recast the condition of QT (\ref{eq_cond_qt_gvm}) in terms of the EPR uncertainty as 
\begin{align}
\det [\mathscr{M}] &= 1 + 4c_{1}c_{2} + ((\eta + \zeta) + 2)((\eta  + \zeta) - (c_{1} + c_{2})) 
\nonumber
\\ 
&~~~~~~~~~~~~~~~~~~~~~~~~~~~~~~~~ - (\eta + \zeta)(c_{1} + c_{2}) < 4
\nonumber
\\
&\Rightarrow \left( (\eta+\zeta) - (c_{1}+c_{2}) + 1 \right)^{2} - (c_{1} - c_{2})^{2} < 4
\nonumber
\\
&\Rightarrow (\eta+\zeta) - (c_{1}+c_{2}) < \sqrt{4 + (c_{1} - c_{2})^{2}} - 1.
\label{eq_cond_tf_epr_gvm}
\end{align}

Evidently, QT ($\det [\mathscr{M}] < 4$) doesn't necessarily imply EPR correlation, in general.
However, considering a restricted form of the Gaussian state (\ref{eq_exp_tmvm_canonical}) where the diagonal entries of the $C$ matrix are equal, i.e., $c_{1} = c_{2} = c$ (let),
we obtain $(\eta+\zeta) - 2c<1$, i.e., for this class of states QT necessarily indicates EPR correlation.
Consequently, this bidirectional between QT and EPR correlation leads to the conclusion that for the limited class of states for which $c_{1}=c_{2}=c$, EPR correlation turns out to be both {\em necessary and sufficient} for QT.

One can further restrict the class of Gaussian entangled states to the case of symmetric Gaussian states for which $\eta=\zeta$.
As quite expected, the condition of EPR correlation (\ref{eq_cond_eprc_gvm}) trivially boils down to $\eta - c < 1/2$.
Similarly, the condition of QT (\ref{eq_cond_qt_gvm}) also simplifies to $(1 + 2(\eta - c))^{2} < 4 \Rightarrow \eta - c <1/2$.
On the other hand, Simon's criterion of inseparability reads $4\Delta - 16\sigma > 1$, where $\Delta = \det[A] + \det[B] - 2\det[C]$ and $\sigma = \det[V]$. 
In the case of symmetric Gaussian states, it is easy to check that Simon's condition reduces to $\left( 4(\eta+c)^{2} - 1 \right)\left( 4(\eta-c)^{2} - 1 \right) < 0$.
From the canonical uncertainty relation we know that $\sqrt{(\eta + c)(\eta - c)} \geq 1/2$.
Considering $\eta + c > 1/2$, we immediately have the Simon's condition for the symmetric Gaussian state reducing to $\eta - c < 1$.
These results lead to the conclusion that for symmetric Gaussian states entanglement alone is both necessary and sufficient for QT \cite{adesso_telport_symgauss}. 
However, in the case of general entangled Gaussian state, entanglement is only a necessary criterion while a stricter condition known as EPR correlation suffices for QT.  

Next, we elaborate our analytical results on the EPR correlation as necessary and sufficient for QT by considering examples of a two-mode squeezed thermal state and the entangled Gaussian resource generated by a passive BS from input single mode nonclassical state.

{\em Two mode Squeezed Thermal State and QT:}
A two mode squeezed thermal state (TMST) is given by, $\rho_{\rm{sth}}^{(2)}=S_{ab}(r)\rho_{\rm{th}}(\bar{n}_{1})\otimes \rho_{\rm{th}}(\bar{n}_{2})S_{ab}^{\dagger}(r)$,
where $S_{ab}(r)=\exp \left[ r(a^{\dagger}b^{\dagger} - ab) \right]$ and $\bar{n}_{i}$ ($i=1,2$) denote the average number of thermal photons standing for the corresponding temperatures.
The covariance matrix for $\rho_{\rm{sth}}^{(2)}$ is given by
\begin{equation}
V_{\rm{sth}}^{(2)}=
\begin{pmatrix}
A & C\\
C^{T} & B
\end{pmatrix}=
\begin{pmatrix}
\eta I & c\sigma_{z} \\
c\sigma_{z} & \zeta I
\end{pmatrix},
\label{eq_exp_vm_tmst}
\end{equation}
where $\eta=\mu^{2}k_{1}+\nu^{2}k_{2}$, $\zeta=\nu^{2}k_{1}+\mu^{2}k_{2}$ and $c=\mu\nu(k_{1}+k_{2})$.
The coefficients $k_{i}(=\bar{n}_{i} + 1/2;i=1,2)$ follow the canonical uncertainty relation $k_{i}\geq 1/2$ ($\because \bar{n}_{i}\geq 0$).
Comparing $V_{\rm{sth}}^{(2)}$ (\ref{eq_exp_vm_tmst}) with the canonical form (\ref{eq_exp_tmvm_canonical}) it is clear that $c_{1} = c_{2}$ leading to the fact for $\rho_{\rm{sth}}^{(2)}$ {\em EPR correlation is both necessary and sufficient to ensure QT}.
We obtain analytic expressions for the conditions of inseparability and EPR correlation/QT for $\rho_{\rm{sth}}^{(2)}$ which are given as
\begin{subequations}
\begin{equation}
r_{\rm{ent}} \geq \frac{1}{2} \ln \left( \frac{1 + 4 k_{1} k_{2} + \sqrt{(4k_{1}^{2} - 1)(4k_{2}^{2} - 1)}}{2(k_{1} + k_{2})} \right)
\end{equation}
\begin{equation}
r_{\rm{epr/qt}} \geq \frac{1}{2} \ln \left( k_{1} + k_{2} \right).
\end{equation}
\label{eq_exp_cond_ent_epr_tmst}
\end{subequations}

\begin{figure}[h]
\includegraphics[scale=0.9]{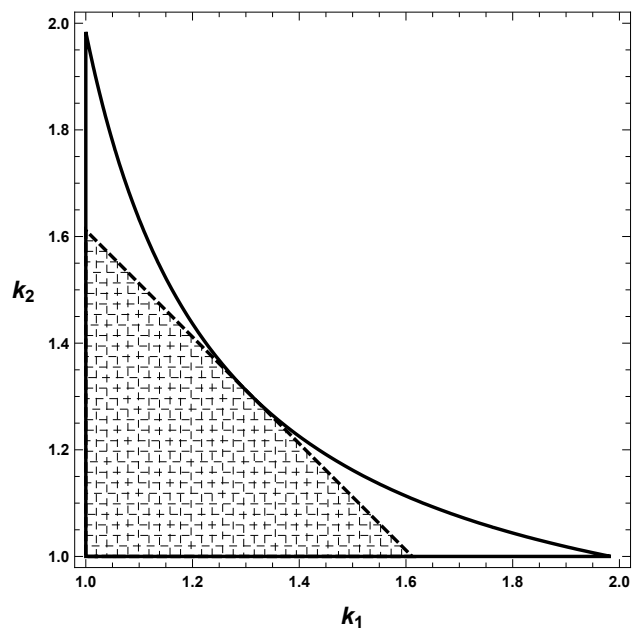}
\caption{Contour plot of the entanglement and EPR correlation in the parameter space for $\rho_{\rm{sth}}^{2}$.
Lower right triangle with dotted line indicates the region of enhanced EPR correlation and the outer polygon represents the region over which the state is entangled.
We consider $r=0.48$.}
\label{fig_ent_epr_tmst}
\end{figure}

Evidently, the condition for EPR correlation is stricter than that for inseparability.
In Fig. \ref{fig_ent_epr_tmst} we plot the regions (space spanned by $k_{1}$ and $k_{2}$) over which TMST is entangled and EPR correlated for the squeeze parameter $r=0.48$.
It is explicit that, for the TMST, the region of entanglement without QT is sufficiently bigger than the region with QT.
Nonetheless, as expected, both the regions match along the line $k_{1}=k_{2}$ where the state becomes symmetric.
It is further observed (not shown in the paper) that with increase in the squeeze parameter $r$ the difference between the {\em without QT} region and {\em with QT} region becomes smaller and asymptotically they merge with each other.
This is quite expected as with increase in $r$ squeezing becomes predominant in TMST.
This, in the asymptotic limit, leads to an effective two-mode squeezed vacuum state, also known as EPR state, which is a symmetric state where the conditions of entanglement and EPR correlation coincide with each other.

{\em BS Generated Entangled Gaussian Resource:}
We consider a BS generated entangled Gaussian resource by sending a single mode squeezed thermal state, $\rho_{\rm{sth}}=S(-r)\rho_{\rm{th}}(\bar{n})S^{\dagger}(-r)$, through one of the input arms of the BS while the other input arm left at vacuum.
The single mode squeezing operator is defined as $S(-r)=\exp \left[ -\frac{r}{2}(a^{\dagger 2} - a^{2}) \right]$ and $\bar{n}$ is the average number of thermal photons representing the temperature of the field.
BS output state is ensured to be entangled by choosing the input state to be nonclassical \cite{nc_bsent_ivan} which could be obtained by considering $r>\frac{1}{2}\ln [2k]$ \cite{nc_sth_asboth, nc_sth_bose}, i.e., making the state quadrature squeezed ($ke^{-2r}<1/2$) \cite{qs_simon}, where $k$ ($=\bar{n}+1/2$) follows the condition $k\geq 1/2$.

Total input covariance matrix is given by $V_{\rm{in}}=\sigma \oplus \frac{I}{2}$ where $\sigma=\rm{diag}\left( k e^{-2r}, k e^{2r} \right)$ and $I$ is $2\times 2$ identity matrix corresponding to the covariance matrix of single mode vacuum state.
Under BS transformation ($U_{\rm{BS}}$) input quadrature vector changes as $U_{\rm{BS}}: \vec{R}_{\rm{in}} \rightarrow \vec{R}_{\rm{out}}=S_{\rm{BS}}\vec{R}_{\rm{in}}$, where $S_{\rm{BS}}=\begin{pmatrix}
\sqrt{T}I & \sqrt{1-T}I\\
-\sqrt{1-T}I & \sqrt{T}I
\end{pmatrix}$. 
Consequently, the output covariance matrix becomes $V_{\rm{out}} = S_{\rm{BS}} V_{\rm{in}} S_{\rm{BS}}^{T}$ and is given by
\begin{equation}
V_{\rm{out}} = \begin{pmatrix}
\eta_{1} & 0 & c_{1} & 0\\
0 & \eta_{2} & 0 & c_{2}\\
c_{1} & 0 & \zeta_{1} & 0\\
0 & c_{2} & 0 & \zeta_{2}
\end{pmatrix} = \begin{pmatrix}
A & C\\
C^{T} & B
\end{pmatrix},
\label{eq_exp_vm_bsgs}
\end{equation}
where $A=T\sigma + (1-T)I/2$, $B=(1-T)\sigma + TI/2$ and $C=\sqrt{T(1-T)}\left( -\sigma + I/2 \right)$.
From the form of $\sigma$ it is clear that $\eta_{1}\leq \eta_{2}$ and $\zeta_{1}\leq \zeta_{2}$.
It also indicates that the sign of $c_{2}$ is negative.
The covariance matrix for the BS generated Gaussian state could be easily brought to the canonical form (\ref{eq_exp_tmvm_canonical}) by suitable local linear unitary Bogoliubov operations (LLUBOs) $U_{L}=U_{L}^{a}\otimes U_{L}^{b}$. 
Since the LLUBOs don't alter the bimodal correlation we work with the current form of the covariance matrix (\ref{eq_exp_vm_bsgs}) to analyze its inseparability, EPR correlation and QT.
Since, we consider the input state is nonclassical ($i.e., r > \frac{1}{2} \ln [2k]$) BS output state is necessarily entangled for all values of $0<T<1$.
However, the input nonclassicality doesn't, in general, guarantee EPR correlation neither QT.
Here, we look at the numerical solution for the EPR correlation and QT.
\begin{figure}[h]
\includegraphics[scale=0.9]{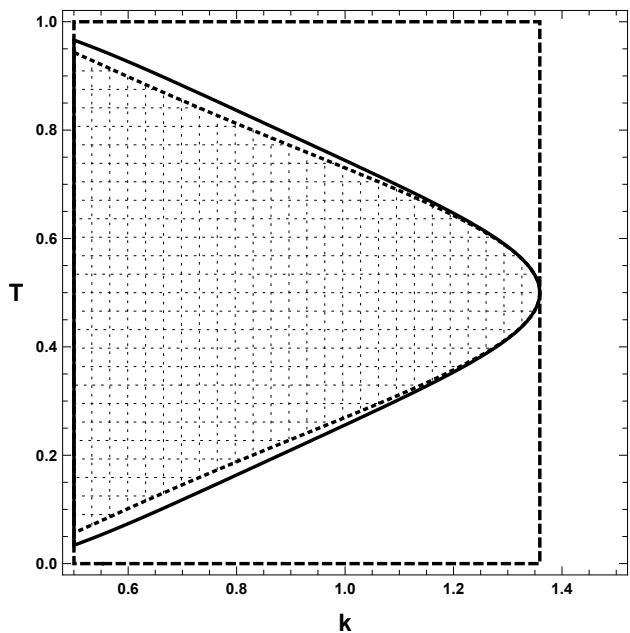}
\caption{Contour plot of entanglement, EPR correlation and QT in the parametric space of $k$ and $T$ for BS output state with input $\rho_{\rm{sth}}$.
The outer rectangular box with dashed boundary represents the region over which the state is entangled.
The region bounded by the solid line stands for the region yielding QT.
Inner dotted region signifies the region where the state is EPR correlated.
We take $r=0.5$.}
\label{fig_ent_epr_smst}
\end{figure}

In Fig. \ref{fig_ent_epr_smst} we plot the parameter regimes over which the BS generated state is entangled, EPR correlated and yields QT.
Evidently, the BS generated Gaussian resource provides a direct example where the parameter region for EPR correlation appears as the subset of that for the QT, i.e., EPR correlation is {\em sufficient for QT; however, not necessary.}
Quite contrary to the earlier example of TMST, here, with increase in the squeeze parameter $r$ the difference between the region of EPR correlation and the region of QT increases (not shown here).
This could be inferred as follows.
As $r$ increases the both the asymmetry in the state and the difference between $|c_{1}|$ and $|c_{2}|$ in the canonical form (\ref{eq_exp_tmvm_canonical}) increase.
As a consequence, it evident that increase in $r$ enhances the gap between EPR correlation and QT.
Nonetheless, in the case of a balanced BS, i.e., $T=1/2$, BS output Gaussian state becomes symmetric for which conditions of both inseparability, EPR correlation and QT become identical as is explicit from the Fig. \ref{fig_ent_epr_smst}.

{\em Discussion:}
In this letter we have analyzed the necessary and sufficient condition (s) for QT with Gaussian resource states.
To that end, we have first proved that although entanglement is necessary, EPR correlation is sufficient for QT, in general.
Moreover, for a restricted class of Gaussian states which are in-between the symmetric states and the most general asymmetric states, in particular the states for which the correlation matrix takes the form $C=\rm{diag}(c,-c)$, EPR correlation turns out to be both necessary and sufficient for QT.
We have also shown that our results immediately boil down to the earlier observation that in the case of symmetric Gaussian states entanglement alone is necessary and sufficient for teleportation beyond classical limit.
In light of the earlier observation on ST \cite{st_grosshans, st_he}, we, in this letter, provide a complete characterization of the resource required for Gaussian teleportation.

We have, then, presented two examples of Gaussian entangled resources, that corroborate our results.
With TMST as entangled Gaussian resource EPR correlation becomes both necessary and sufficient for QT.
However, in the case of BS generated entangled Gaussian resource that belongs to the class of most general asymmetric Gaussian states, EPR correlation is only sufficient for QT.
It is noteworthy that the case of BS generated resource is generated by using nonclassical input at one of the input arms.
However, it could be easily generalized to the case of input at both the arms. 
In the latter case, the canonical form of the variance matrix, we have considered in this letter, is generated by setting the individual squeezing parameters different.

Our results could be easily extended to a more realistic situation where the input state is chosen from a preselected input set.
This renders relevance to our primary result in the context of various quantum benchmarks \cite{benchmark_hammerer, benchmark_adesso, benchmark_owari, benchmark_calsamiglia, benchmark_chiribella} of continuous variable teleportation.
Moreover, with prior knowledge of the input state, our sufficient and necessary criteria immediately becomes applicable to the remote state preparation scenario \cite{rsp_kurucz, rsp_pogorzalek}.
In view of the recent advances in the quantum technology science \cite{Pirandola_NatPhys}, our result could be extended to other continuous variable systems such as opto-mechanical systems \cite{cvsystem_opm}.

Author expresses his gratitude to M. Sanjay Kumar, S. N. Bose National Centre for Basic Sciences, kolkata for numerous discussion on the sufficient criterion for QT with Gaussian resources.
Author is also indebted to R. Simon, IMSc, Chennai for pointing out BS generated resources as example of states having most general asymmetric variance matrix.


\end{document}